\documentclass[useAMS,usenatbib]{mn2e}
\usepackage{epsfig}
\usepackage{amsmath}
\usepackage{rotating}

\newcommand{\be}{\begin{equation}}
\newcommand{\beq}{\begin{equation}}
\newcommand{\ba}{\begin{eqnarray}}
\newcommand{\ee}{\end{equation}}
\newcommand{\eeq}{\end{equation}}
\newcommand{\ea}{\end{eqnarray}}

\newcommand{\hs}{\hspace{1mm}}

\newcommand{\lya}{Ly$\alpha \hspace{1mm}$}

\def\lsim{~\rlap{$<$}{\lower 1.0ex\hbox{$\sim$}}}

\def\gsim{~\rlap{$>$}{\lower 1.0ex\hbox{$\sim$}}}

\title[Combined constraints on intergalactic dust]{Combined constraints on intergalactic dust from quasar colours and the soft X-ray background}
\author[J. Johansson et al.]{
Joel Johansson$^{1,2}$\thanks{E-mail:joeljo@fysik.su.se}
and
Edvard M\"ortsell$^{1,2}$ \\
$^{1}$Physics Department, Stockholm University, AlbaNova University Center, SE--106 91,
Stockholm, Sweden \\
$^{2}$The Oskar Klein Center, Stockholm University, SE--106 91,
Stockholm, Sweden}
\begin{document}

\label{firstpage}

\maketitle

\begin{abstract}
Unless properly corrected for, the existence of intergalactic dust will introduce a redshift dependent magnitude offset to standard candle sources. This would lead to overestimated luminosity distances compared to a dust-free universe and bias the cosmological parameter estimation as derived from, e.g., Type Ia supernovae observations. 
In this paper, we model the optical extinction and X-ray scattering properties of intergalactic dust grains to constrain the intergalactic opacity using a combined analysis of observed quasar colours and the soft X-ray background.
Quasar colours effectively constrain the amount of intergalactic dust grains smaller than $\sim 0.2 \, \mu$m, to the point where we expect the corresponding systematic error in the Type Ia supernova magnitude-redshift relation to be sub-dominant. Soft X-ray background observations are helpful in improving the constraints on very large dust grains for which the amount of optical reddening is very small and therefore is more difficult to correct for. Our current upper limit corresponds to $\sim 0.25$ magnitude dimming at optical wavelengths for a source at redshift $z=1$, which is too small to alleviate the need for dark energy but large in terms of relative error. 
However, we expect it to be possible to lower this bound considerably with an improved understanding of the possible sources of the X-ray background, in combination with observations of compact X-ray sources such as Active Galactic Nuclei. 
\end{abstract}


\section{Introduction}
The use of Type Ia supernovae (SNe Ia) as standardized candles to probe the redshift-distance relation remains essential for establishing and exploring the dark energy universe. 
The fact that complementary cosmological probes agree on the concordance cosmological model with current accelerated cosmological expansion makes it unlikely that the observed dimming of SNe Ia are solely due to dust extinction. Nonetheless, it is still possible that dust extinction could bias cosmological parameter estimation. Also, when trying to explore specific dark energy properties, it becomes crucial to be able to constrain the effect of dust on SN Ia observations \citep{menard2010b, corasaniti2007}.

The presence of dust in the intergalactic medium (IGM) has been the subject of numerous studies.
Based on estimates of the stellar density and metallicity as a function of redshift, several authors have inferred the existence of significant amounts of dust in the IGM with density $\Omega_{\rm dust} \sim 10^{-6} - 10^{-5}$ \citep{loebhaiman1997,inoue2004, fukugita2011}.

Dust grains scatter and absorb photons with an energy dependent cross-section and typical dust extinction is correlated with reddening of the incoming light.
The amount of reddening is usually quantified by the total-to-selective extinction ratio $R_V \equiv A_V/E(B-V)=A_V/(A_B-A_V)$, where $E(B-V)$ is the colour excess and $A_B$ and $A_V$ are the extinction in the $B$ and $V$-band respectively.
\citet{menard2010a} report a statistical detection of dust reddening of quasars (QSOs) out to large distances (a few Mpc) around galaxies at $z = 0.3$. The observed reddening implies a slope of the extinction curve, $R_V = 3.9 \pm 2.6$, which is consistent with that of Milky Way dust ($R_V = 3.1$), albeit with large uncertainties. Extrapolating this result to higher redshifts yield a lower limit estimate of the extinction in the restframe $B$-band of $A_{B}(z=1) \gsim 0.03$ magnitudes. 

\citet{aguirre1999b} and \citet{bianchi2005} find that astrophysical processes which transfer dust into the intergalactic medium would preferentially destroy small grains, leaving only grains larger than $a \sim 0.1 \mu$m, implying less reddening and higher values for $R_V$ (often labeled ``grey" dust). The absence of detectable systematic reddening of SNe Ia with increasing redshift implies that any intergalactic dust extinction must be quite grey at optical wavelengths, suggesting the grains must be large. 
Since the dust correction for SNe Ia depends on the reddening, large grains will more difficult to correct for.
\citet{mortsell2003} and \citet{mortsell2005} simulated the reddening by intergalactic dust based on the different parametrizations of mean extinction laws of Milky Way-like dust, with $0 < R_{V} < 12$. They used observations of QSO colours and template spectra to put an upper limit on the dimming in the restframe $B$-band by intergalactic dust of a source at redshift $z=1$ of $A_{B}(z=1)\lsim 0.02$ for $R_V\sim 3$ and $A_{B}(z=1) \lsim 0.1$ for $R_V \sim 10$. 

Apart from the reddening of cosmological sources, the presence of dust grains can also be inferred from the absorption of UV/optical photons which are re-emitted in the far-infrared.
\citet{aguirrehaiman2000} calculated the contribution from intergalactic dust to the cosmic far-infrared background, assuming it was heated by intergalactic radiation. 
They found that dust densities of $\Omega_{\rm dust} \sim {\rm a \, few} \times 10^{-5}$, necessary to account for the dimming of SNe Ia, would produce most of the far-infrared background.
However, observational data leaves little room for any such diffuse emission component, since discrete sources detected by the SCUBA survey account for almost all of the background at 850 $\mu$m \citep{hauserdwek2001}.

Intergalactic dust would also scatter X-rays from point sources into extended, diffuse X-ray halos.
If the size of the halo is large enough, the scattered radiation will effectively be part of the unresolved soft X-ray background.
\citet{dl2009} argue that dust scattered X-ray halos around Active Galactic Nuclei (AGN) can maximally account for a fraction $f_{\rm halo} \sim 5-15\%$ of the total measured Soft X-ray Background (SXB). This allows them to place an upper limit on the optical/near-infrared extinction $\Delta m(z=1, \lambda=8269\mbox{\AA})\lsim 0.15 (f_{\rm halo}/10\%)$.
The absence of an X-ray halo around a single $z = 4.3$ QSO observed with {\it Chandra}, allow \citet{petric2006} and \citet{corrales2012} to place upper limits on the dust density parameter $\Omega_{\rm dust} \lsim Ê10^{-6} - 10^{-5}$ assuming a constant comoving number density of dust grains of size $a = 1 \,\mu$m or with a power law distribution of grain sizes $0.1\leq a \leq 1\,\mu$m.

In this paper we will combine a QSO colour analysis with a SXB analysis to constrain the amount of both small and large dust grains. The outline of the paper is as follows: In Section 2 we continue discussing existing constraints on intergalactic dust and argue for a physical model of the intergalactic dust. In Section 3 we evaluate the effects of intergalactic dust on QSO colours and in Section 4 we examine how the unresolved SXB can put further constraints on these dust models. We present our results in Section 5 and summarize and discuss these results and some future prospects in Section 6.
\vspace{-12pt}
\section{Intergalactic dust} 
Dust grains are present in our Galaxy, in the host galaxies of cosmological sources, in galaxies along the line-of-sight and possibly in the immediate surroundings of the source \citep[e.g.][]{schlegel1998,ostman2006,ostman2008,goobar2008}. In this paper, we will study the effect of intergalactic dust along the line-of-sight to cosmological sources. 
Extinction by interstellar dust in the Milky Way and nearby galaxies, such as the Small and Large Magellanic Clouds (SMC and LMC), is usually parameterized using mean extinction laws, $A_\lambda = f(\lambda,R_V)$ where $A_\lambda$ is the extinction for wavelength $\lambda$ and the reddening parameter, $R_{V}= A_{V}/(A_{B} - A_{V})$, is the slope of the extinction curve in the optical region \citep[e.g.][]{ccm1989,fitzpatrick1999}.  Within the Milky Way, the reddening parameter ranges between $2 \lsim \, R_{V} \lsim 6$ for different sightlines and is usually approximated as $R_{V}=3.1$. It is not known if these parametrizations are valid for an intergalactic dust population where the dust properties might be different from those in the interstellar medium. Small grains preferentially scatter light with short wavelengths, producing a steep extinction law with small values for $R_V$. Very large grains would produce wavelength-independent grey extinction with $R_V \rightarrow \infty$. 

To build up extinction curves from a generic population of dust grains one needs to know the composition, size distribution, scattering and absorption properties of the dust grains. 
\citet{mrn1977}, hereafter MRN, found that interstellar extinction in the Milky Way is well fitted using separate populations of bare silicate and graphite grains with a power-law distribution of sizes where $dn/da \propto a^{-3.5}$. Within the Milky Way, graphite grains typically range in size from 0.005 to 1 $\mu$m while silicate grain sizes range from 0.025 to 0.25 $\mu$m. 

We will study dust models that have either a single grain size or truncated MRN size distributions, varying the size of the smallest and largest grains $a_{\rm min}$ and $a_{\rm max}$. The intergalactic dust is assumed to be distributed homogeneously with a constant comoving number density of dust grains, $n(z)$. It is reasonable to assume that the dust density traces the stellar mass density in the universe, and we also investigate cases where the comoving dust density is proportional to the integrated star formation rate density, $n(z) \propto \int_{z}^{\infty} dz' \frac{\dot{\rho}_{\star}}{(1+z')\mathcal{E}(z')}$, where we use the analytical expression for the star formation rate, $\dot{\rho}_{\star}$, derived by \citet{hernquistspringel2003}.
\subsection{Optical depth to scattering and absorption}
The total optical depth to scattering and/or absorption by dust grains between redshift 0 and $z_{\rm em}$ can be calculated as
\begin{equation}
\tau \left( \lambda_{\rm obs}, z_{\rm em} \right) = \frac{c}{H_{0}} \int_{0}^{z_{\rm em}} dz' \frac{n(z') \sigma (\lambda')(1+z')^{2}}{\mathcal{E}(z')} \, ,
\label{eq:tau_fixed}
\end{equation}
where $n(z')$ is the comoving number density of dust grains at redshift $z'$. Each dust grain is assumed to be spherical, with radius $a$ and geometrical cross section $\sigma(\lambda')= \pi a^{2} Q(\lambda')$ where the efficiencies for scattering and absorption are $Q_{\rm scat}$ and $Q_{\rm abs}$, the efficiency for extinction is $Q_{\rm ext}=Q_{\rm scat}+Q_{\rm abs}$, and $\lambda' = \lambda_{\rm obs} / (1+z)$. The dimensionless expansion factor $\mathcal{E}(z') = \sqrt{\Omega_{M}(1+z)^{3} +\Omega_{\Lambda}}$. For a distribution of grain sizes, Eq. \ref{eq:tau_fixed} becomes
\begin{equation}
\tau \left( \lambda_{\rm obs}, z_{\rm em}\right) = \frac{c}{H_{0}} \int_{0}^{z_{\rm em}} dz' \int_{a_{\rm min}}^{a_{\rm max}} da \frac{dn}{da} \frac{ \sigma (\lambda')(1+z')^{2}}{\mathcal{E}(z')} \, ,
\label{eq:tau_distr}
\end{equation}
where $\frac{dn}{da} da$ denotes the comoving number density of dust grains with radii in the range $a \pm da/2$.

\begin{figure}
\includegraphics[width=84mm]{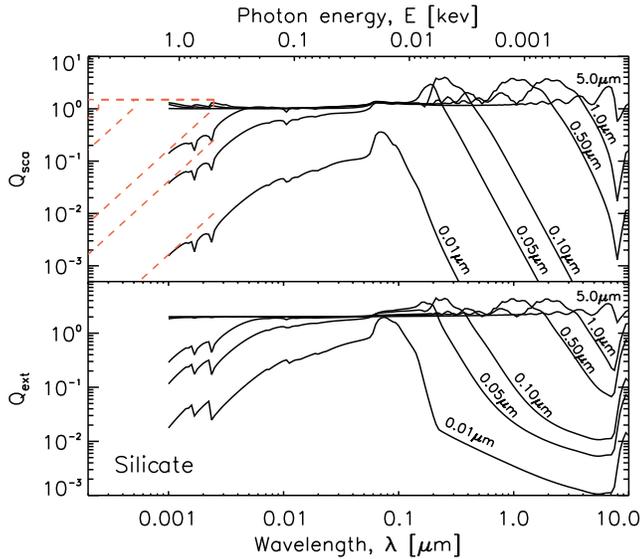}
\caption{Upper panel: Scattering efficiency as function of wavelength, $Q_{\rm sca}(\lambda)$ for spherical silicate dust grains of radii, $a = 0.01 - 5.0\mu$m. Solid black lines show $Q_{\rm sca} = \sigma_{\rm sca} / \pi a^2$, where $\sigma_{\rm scat}$ is the scattering cross section, from \citet{laordraine1993} and the dashed red lines show the approximation made in Eq. \ref{eq:Q}. Lower panel:  Extinction efficiency as function of wavelength, $Q_{\rm ext}(\lambda) = Q_{\rm sca} + Q_{\rm abs}$ for spherical silicate dust grains of radii, $a = 0.01 - 5.0\mu$m. }
\label{q_suvsil}
\end{figure}

In this paper we focus on dust composed of silicate ($\rho_{\rm grain} \approx 3.2$ g/cm$^{3}$) and graphite grains ($\rho_{\rm grain} \approx 2.3$ g/cm$^{3}$) \citep[described in][]{drainelee1984,laordraine1993,weingartnerdraine2001}. The original astronomical silicate model constructed by \citet{drainelee1984} used laboratory measurements of crystalline olivine in the vacuum ultraviolet, resulting in an absorption feature at $\lambda \sim 0.15 \,\mu$m not seen in astronomical objects. In this paper, we use the smoothed astronomical silicate model dielectric function obtained by removing this absorption feature \citep{weingartnerdraine2001}.
The publicly available scattering and absorption efficiencies $Q_{\rm sca}$ and $Q_{\rm abs}$ are given for wavelengths $\lambda=10^{-3}$ to $10^{3}$ $\mu$m for grains with radii $a=10^{-3}$ to 10 $\mu$m, see Fig. \ref{q_suvsil}. For wavelengths shorter than $\lambda \lsim 10^{-3}$ $\mu$m (corresponding to X-ray energies $E \gsim 0.5$ keV) we use the scattering approximations  \citep[]{alcockhatchett1978, miralda1999, dl2009}
\begin{equation}
\begin{array}{lllc} 
	Q_{\rm scat} & \approx  & \left\{ \begin{array}{l} 0.7(a / \mu{\rm m})^2 (6 \hs {\rm keV}/E)^2 \\  1.5 \end{array} \right. & \begin{array}{l}\mbox{$Q_{\rm scat} < 1$}\\ 
\mbox{otherwise.}\end{array} 
\end{array} 
\label{eq:Q}
\end{equation} 

\begin{figure}
\includegraphics[width=84mm]{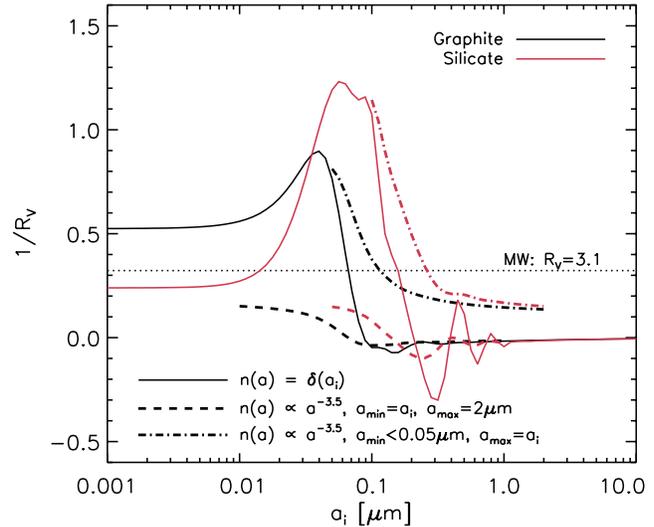}
\caption{$R_{V}^{-1} = (A_B - A_V)/A_V$ reddening for dust models with graphite (black lines) and silicate (red lines) grains with a single grain size (solid lines) or models with truncated MRN grain size distributions with $0.05\mu$m $<a_{\rm min} < 1.0\mu$m and $a_{\rm max} = 2.0\, \mu$m (dashed lines) or $a_{\rm min} = 0.02\,\mu$m (graphite) or $a_{\rm min} = 0.05\, \mu$m (silicate) and $0.1 < a_{\rm max} < 2.0\, \mu$m (dotted-dashed lines). The inclusion of grains with radii larger than $a \gsim 0.1-0.2 \mu$m gives gray extinction with $R_{V}^{-1} \rightarrow 0$. The horizontal dotted line shows the average reddening for interstellar dust in the Milky Way, $R_{V}=3.1$}
\label{rv}
\end{figure}

In Fig. \ref{rv}, we show the corresponding total-to-selective extinction ratio $R_V=A_V/(A_B-A_V)$ for the graphite (black lines) and silicate dust (red lines) models employed in this paper. Solid lines correspond to single size models, dashed and dash-dotted lines to truncated MRN grain size distributions where $a_{\rm min}$ and $a_{\rm max}$ are varied, respectively. The inclusion of large grains gives $A_B \approx A_V$, which drives $R_{V}^{-1} \rightarrow 0$. Note that the $R_V$-value only refers to the modelled optical extinction in the $B$- and $V$-band and does not necessarily describe the entire extinction curve as observed for Milky Way dust. In fact, the variations in the $Q$-values for silicate dust grains $a \sim 0.25\, \mu$m can also introduce a small blueing-effect with $R_V \lsim 0$.

\subsection{X-ray scattering by dust}
Small angle scattering of X-rays by dust grains along the line-of-sight can produce diffuse halos around X-ray point sources, as observed for many galactic X-ray sources. Measurements of the intensity and angular extent of such halos provide a quantitative test of interstellar grain models \citep{mg1986, ml1991, dwek1998,draine2003}.
In a similar way, intergalactic dust grains would produce diffuse halos around X-ray sources on cosmological distances. In common with X-ray halos seen around galactic point sources, the angular extent of an intergalactic halo is determined by the scattering properties of the grains \citep[]{evans1985, vaughan2006, petric2006}.

The differential cross sections in the Rayleigh-Gans approximation for a spherical dust grain of radius $a$ is given by \citep{hayakawa1970,mg1986}
\begin{equation}
\frac{d\sigma_{sca}}{d\Omega} = A_E \left(\frac{a}{1.0 \mu \mbox{m}}\right)^{6} \left[ \frac{j_1(x)}{x}\right]^2 (1 + \cos^2 \theta) \, ,
\label{rayleighgans}
\end{equation}
where $x=(4\pi a/\lambda)\sin(\theta/2)$ and $j_{1}(x)=(\sin x)/x^{2} - (\cos x / x)$ is the spherical Bessel function of the first order. $A_{E}$ is a normalization depending on the grain composition and energy $E$ of the X-rays being scattered,
\begin{equation}
A_{E} = 1.1 \left( \frac{2Z}{M} \right)^{2} \left( \frac{\rho_{\rm grain}}{3 \mbox{ g cm}^{-3}} \right)^{2} \left( \frac{F(E)}{Z} \right)^{2} \, ,
\end{equation}
where $F(E)$ is the atomic scattering factor \citep{henke1982, henke1993}, $Z$ is the atomic charge, $M$ is the atomic mass number and $\rho_{\rm grain}$ is the mass density of the dust grain.
The central core of Eq. \ref{rayleighgans} is approximately gaussian, and the root mean square (rms) of the scattering angle $\theta$ indicates the typical size of the scattered halo \citep{mg1986},  $\theta_{\rm rms} \approx 62.4 \left(1.0\, \mu{\rm m}/ a \right) \left(1{\rm \, keV}/ E \right)$ arcsec. 

For low optical depths ($\tau_{\rm sca} \ll 1$), the halo fluence per unit scattering angle is \citep{vaughan2006},
\begin{equation}
\label{halofluence}
\frac{dF_{\rm halo}}{d\theta} \propto 2\pi \theta A_E \left(\frac{a}{1.0 \mu \mbox{m}}\right)^{6} \left[ \frac{j_1(x)}{x}\right]^2 (1 + \cos^2 \theta) N_{g} F_{X} \, ,
\end{equation}
where $N_{g}$ is the dust grain column density along the line-of-sight and $F_{X}$ is the unscattered flux of the X-ray point source.

\citet{dwek1998} showed that the Rayleigh-Gans approximation and the more exact Mie theory are in close agreement, given that the photon energy $E$ (in keV) of the X-ray being scattered is larger than the grain radius $a$ (in $\mu$m).
For our purposes, where we want to study the contribution of dust scattered X-ray halos around AGN to the 0.5-2.0 keV Soft X-ray Background, we note that the average photon energy at the site of the scattering dust grain is higher than the observed energy by a factor $(1+z_{\rm grain})$, which tends to alleviate the error introduced by the use of the Rayleigh-Gans approximation for the larger grains. 

\section{Constraints from quasar colours}
QSOs have been found to be relatively homogeneous in terms of colours and spectral features over a large redshift range. The SDSS DR7 Quasar Catalogue \citep{schneider2010} contains 105783 spectroscopically confirmed QSOs with optical magnitudes measured through five broadband filters ($u,g,r,i,z$). 
We only include point sources (removing QSOs flagged as extended) brighter than the limiting magnitudes $[u,g,r,i,z] = [22.3,22.6,22.7,22.4,20.5]$, corresponding to $S/N$ greater than 5:1. Objects deviating by more than $2\sigma$ from the mean colour are rejected \citep{mortsell2003,mortsell2005}.

\subsection{Composite quasar spectral template}
\begin{figure}
\includegraphics[width=84mm]{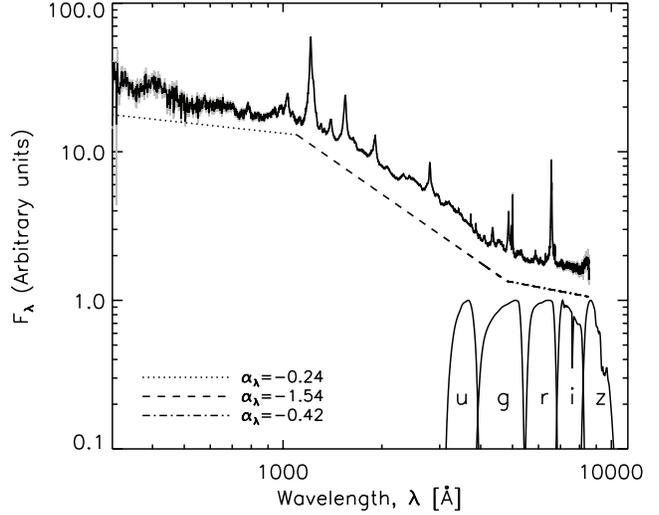}
\caption{The composite QSO spectral template together with a broken power law, $F_{\lambda} \propto \lambda^{\alpha_{\lambda}}$, with spectral indices  $\alpha_{\lambda} = -0.24$ (dotted line, for $\lambda < 1200$\AA), $\alpha_{\lambda} = -1.54$ (dashed line, for $1200 < \lambda < 4850$ \AA) and $\alpha_{\lambda} = -0.42$ (dotted-dashed lines for $\lambda > 4850$\AA ). Shown are also the SDSS $ugriz$ transmission curves.}
\label{qsotemplate}
\end{figure}

For our purposes it is preferable to use an unabsorbed QSO spectrum. We use the SDSS median composite spectrum \citep[as derived by][]{vandenberk2001} spliced with the HST radio-quiet composite spectrum \citep{telfer2002} to achieve a wavelength range $302 < \lambda < 8552$ \AA.
The continuum is well fitted by a broken power law ($F \propto \lambda^{\alpha_\lambda}$) with spectral indices $\alpha_{\lambda} = -0.24$ blueward of Ly$\alpha$ ($\lambda \lsim 1200$ \AA$), \alpha_{\lambda} = -1.56$ for $1200 < \lambda < 4850$ \AA \, and $\alpha_{\lambda} = -0.42$ for $\lambda > 4850$ \AA, see Fig. \ref{qsotemplate}. Although the statistical uncertainties in the continuum fits are small, the quoted systematic uncertainties in the spectral indices are $\sigma_{\alpha_{\lambda}} \sim 0.10 - 0.15$.

\subsection{Simulated Quasar Colours}
The observed dust attenuated flux $F_{\rm obs}$ of an object at redshift $z_{\rm em}$ observed at wavelength $\lambda_{\rm obs}$ is given by
\begin{equation}
F_{\rm obs} (\lambda_{\rm obs}, z_{\rm em}) = F_{\rm em} \cdot e^{-{\tau_{\rm ext} \left( \lambda_{\rm obs}, z_{\rm em} \right)}} \, ,
\label{eq:flux}
\end{equation}
where $F_{\rm em}$ is the intrinsic flux and the optical depth to extinction $\tau_{\rm ext}$ can be calculated using Eq. \ref{eq:tau_fixed} and \ref{eq:tau_distr} for graphite and silicate dust grains. 
The dimming (i.e. increase in apparent magnitude) of light emitted at redshift $z_{\rm em}$ observed at a wavelength $\lambda_{\rm obs}$ is then $\Delta m \left( \lambda_{\rm obs}, z_{\rm em} \right) = \frac{2.5}{\ln 10} \tau_{\rm ext} \left( \lambda_{\rm obs}, z_{\rm em} \right)$. 
For each dust model we simulate the attenuation of the median QSO template spectrum and perform synthetic photometry using the SDSS $ugriz$-filter functions. For each pair of filters $X$ and $Y$, we compare the simulated colours with observed colours,
\begin{equation}
\label{deltaxy}
\Delta (X-Y)_{z} \equiv (X-Y)_{z}^{\rm obs} - (X-Y)_{z}^{\rm sim}.
\end{equation}
For each set of filters $X$ and $Y$, we calculate the mean colour $(X-Y)_{z}^{\rm obs}$ in redshift bins of $\Delta z = 0.05$ in the redshift range $0.5 < z < 3$. At higher redshifts the flux decrement due to the \lya break makes the bluer bands inefficient, so for the $u$- and $g$-band we only include QSOs with $z<1.5$ and $z<2.3$ respectively in the analysis. 

The probabilities of different dust scenarios for each colour $X-Y$ are then calculated using
\begin{equation}
\label{chi2xy}
\chi^{2}(X-Y) = \sum_{i,j=1}^{N} \Delta (X-Y)_{z_i} V(X-Y)^{-1}_{j,i} \Delta (X-Y)_{z_j}\, ,
\end{equation}
where $i$ and $j$ index the $N$ redshift bins. The covariance matrix $V(X-Y)_{i,j}$ is
\begin{eqnarray}
V(X-Y)_{i,j} &=& \sigma(X-Y)^{2}_{\rm cal} + \\ \nonumber
 & &  \sigma(X-Y)^{2}_{\rm temp} + \delta_{ij} \sigma(X-Y)^{2}_{\rm obs}.
\end{eqnarray}
The uncertainties in the photometric calibration $\sigma(X-Y)_{\rm cal} \lsim 0.02$ and the typical observed colour variance in the QSO sample $\sigma(X-Y)_{\rm obs}^{2} \lsim 0.003$ for redshifts $0.5 < z < 2.5$. 
The major source of uncertainty in this analysis is $\sigma(X-Y)_{\rm temp}$, due to the tilting of the spectral template according to the systematic uncertainties of the measured continuum spectral indices, $\sigma_{\alpha_{\lambda}} \approx 0.10$. In practice, we combine the results from all possible colour combinations taking into account the correlations between different colours and redshifts. This is accomplished using Monte Carlo simulations to create a total covariance matrix. 

\section{Constraints from the Soft X-ray Background}
\subsection{The Cosmic X-ray Background}
The cosmic X-ray background comprises the integrated emission of X-ray sources, primarily AGN, extended emission from galaxy clusters, faint starburst and ``normal" galaxies.
\citet{moretti2003} measure the total Soft X-ray Background (SXB) in the 0.5 - 2 keV band to be $(7.53Ê\pm 0.35)Ê\times 10^{-12}$ ergs/s/cm$^2$/deg$^2$ and find that $94_{-6.7}^{+7.0}$\% of the SXB can be ascribed to discrete source emission, i.e $\sim 6 \pm 6$\% of the SXB is unresolved. Similarly, \citet{hm2007a} find that the total SXB in the 1-2 keV band is $(4.6Ê\pm 0.3)Ê\times 10^{-12}$ ergs/s/cm$^2$/deg$^2$. After excluding regions of radius $\lsim 2-20"$ around detected X-ray, {\it HST} and {\it Spitzer} sources, they find a remaining unresolved SXB of $(3.4Ê\pm 1.4)Ê\times 10^{-13}$ ergs/s/cm$^2$/deg$^2$, which is 7.3$\pm$3.0\% of the total SXB. 

\subsection{AGN X-rays scattered by dust}
AGN produce a dominant fraction ($\sim 80\%$) of the SXB at energies $0.5 < E < 2.0$ keV. If dust pervades throughout the intergalactic medium, its scattering opacity would produce diffuse X-ray halos around AGN. If the angular extent of these halos are large enough, they will effectively contribute to the unresolved SXB. This fact, along with the observational upper limits on the unresolved SXB, allows us to constrain the amount and properties of intergalactic dust.

\citet{dl2009} argue that dust scattered X-ray halos around AGN can maximally account for a fraction $f_{\rm halo} \sim 5-15\%$ of the SXB, allowing them to put upper limits on the opacity of intergalactic dust grains $\tau_{\rm dust}(z=1, \lambda=8269\mbox{\AA})\lsim 0.1 - 0.2 (f_{\rm halo}/10\%)$, depending on grain size or size distribution. In the spirit of this approach, we calculate the flux of X-ray photons observed in the Soft X-ray energy range $E = 0.5 - 2.0$ keV, which is expected to be scattered into halos,
\begin{equation}
F_{\rm halo} = \int_{0.5{\rm \, kev}}^{2.0{\rm \, kev}} S(E) dE  \int_{0}^{\infty} \mathcal{F}(z') \times \left[e^{\tau_{\rm sca}(E,z')} - 1\right] dz' \, ,
\end{equation}
where the optical depth to scattering $\tau_{\rm sca}$ is calculated using Eq. \ref{eq:tau_fixed} and \ref{eq:tau_distr}. We take the observed spectral energy density of the AGN to be given by $S(E) \propto E^{-1.4}$ normalized to unity over the energy range \citep[the final results of the analysis depend only weakly on the choice of the spectral index, ][]{moretti2003,dl2009} and,
\begin{equation}
\mathcal{F}(z) = \frac{\mathcal{L}(z)}{(1+z)^{2} \mathcal{E}(z)}.
\end{equation}
The comoving X-ray emissivity $\mathcal{L}(z)$ of AGN is expressed in terms of an integral over the AGN luminosity function
\begin{equation}
\mathcal{L}(z) = \int_{L_{\rm min}}^{L_{\rm max}} L \times \psi(L,z) d \log L .
\end{equation}
The X-ray luminosity function $\psi(L,z) d \log L$ is described by the fitting formula of \citet{hopkins2007} with $\log L_{\rm min} = 40.4$ and $\log L_{\rm max} = 48.0$. The total contribution to the SXB in the observed 0.5-2.0 keV band is obtained by integrating the X-ray emissivity of AGN over redshift, $F_{\rm AGN} = \int_{0}^{\infty} \mathcal{F}(z') dz' = 5.74\times10^{-12}$ ergs/s/cm$^2$/deg$^2$.\\ 

In the following, we conservatively argue that the amount of flux from AGN being scattered far from the original line of sight, i.e. into an extended halo, $F_{\rm halo}$, will effectively be measured as a diffuse background and can thus not be larger than the measured value of the unresolved SXB, $F_{\rm uSXB}$.

\begin{figure}
\includegraphics[width=84mm]{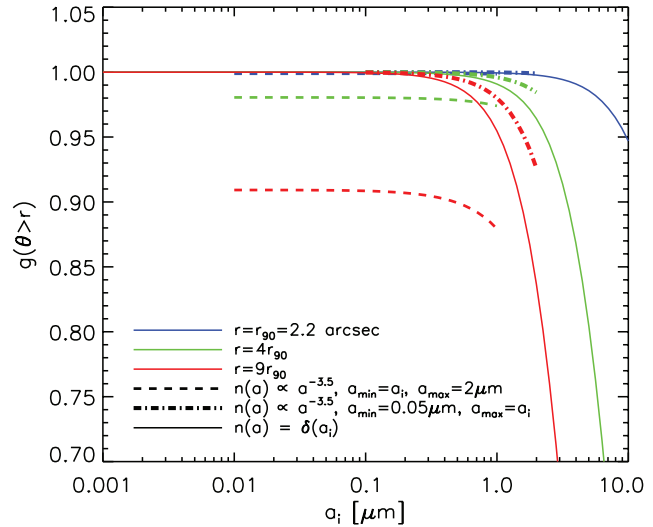}
\caption{Fraction of halo flux falling outside an aperture of radius $r=r_{90}=2.2"$ (blue lines), $r = 4r_{90}$ (green lines), $r = 9r_{90}$ (red lines) calculated using Eq. (\ref{halofluence}) integrated over angles and an energy range between $E_{\rm min}=0.5$ keV and $E_{\rm max}=2.0$ keV assuming a spectral energy distribution, $S(E) \propto E^{-1}$. Solid lines show results for single grain sizes, dashed lines show models with truncated MRN grain size distributions with $0.01 <a_{\rm min} < 1.0\, \mu$m and $a_{\rm max} = 2.0\, \mu$m and dotted-dashed lines show models with truncated MRN grain size distributions with $a_{\rm min} = 0.05\, \mu$m and $0.1 < a_{\rm max} < 2.0\, \mu$m.}
\label{halofractions}
\end{figure}

When \citet{hm2007a} measure the unresolved part of the SXB, they exclude regions of radius $\theta < r_{90}=2.2$ arcsec for {\it HST} and {\it Spitzer} sources undetected in X-rays and regions $\theta < 4-9 r_{90}$ for X-ray detected sources. Dust grains having $a \gsim 2\, \mu$m will scatter X-rays into very compact halos with angular sizes comparable to these excluded regions, thus contributing less flux to the unresolved SXB. While the scattering is dominated by the largest grains, inclusion of smaller grains would cause the halo flux to fall off less rapidly at large angles, and the resulted tail could contain a substantial flux which would be effectively be considered as an unresolved background. We verify that a major fraction, $g(\theta > r)$, of the dust scattered halo flux falls outside these apertures (see Fig. \ref{halofractions}) for dust models with single grain sizes or MRN grain size distributions, and will henceforth assume that similar exclusion regions apply to the SXB measurements of \citet{moretti2003}.

We calculate the probability of different dust scenarios using
\begin{equation}
\chi_{\rm SXB}^{2} = \frac{\left( F_{\rm uSXB}^{\rm obs} - F_{\rm halo}^{\rm mod} \right)^2}{\sigma_{\rm obs}^2+\sigma_{\rm mod}^2} \, ,
\end{equation}
where $F_{\rm uSXB}^{\rm obs}$ is the unresolved Soft X-ray Background in the 0.5-2 keV band as calibrated by \citet{moretti2003} and $F_{\rm halo}^{\rm mod} = F_{\rm halo} \cdot g(\theta > r)$  is the modelled halo flux falling outside an aperture of radius $r$. For values $F_{\rm halo}^{\rm mod}\leq F_{\rm uSXB}^{\rm obs}$, we set the corresponding $\chi^{2}$-value to zero. The model uncertainty will be negligible compared to the uncertainty of the unresolved SXB $\sigma_{\rm obs} = 4.52 \times 10^{-13}$ ergs/s/cm$^2$/deg$^2$.
\vspace{-12pt}
\section{Results}
We present our results from the combined analysis of QSO colours and the unresolved SXB together with the expected dimming in the restframe $B$- and $V$-band for a source at redshift $z=1$, $A_{B}(z=1)$ and $A_{V}(z=1)$. We also express our constraints as upper limits on the density parameter,  $\Omega_{\rm dust} = \left(n \frac{4\pi}{3} a^{3} \rho_{\rm grain}\right) / \rho_{\rm crit}$, where $\rho_{\rm crit} = 1.88Ê\times 10^{-29} h^{2}$ g cm$^{-3}$.
\subsection{Dust models with single grain sizes}
\begin{figure}
\includegraphics[width=84mm]{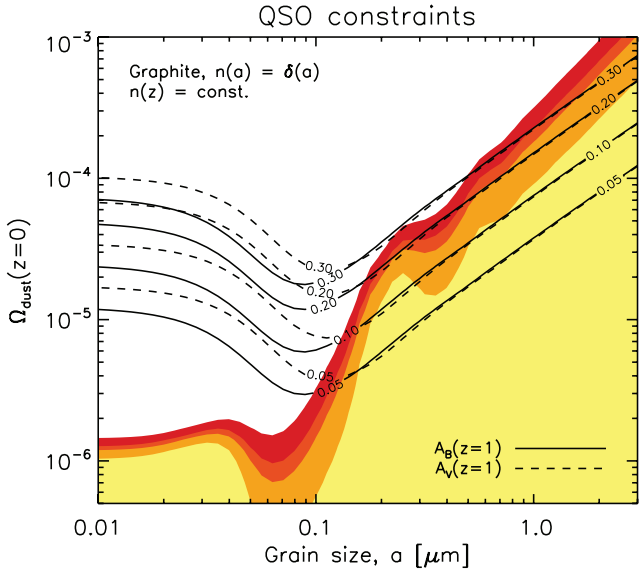}
\caption{Constraints from the QSO colour analysis on dust models with single size graphite grains and a constant comoving number density. Regions (from yellow to red) indicate allowed dust models at 68\%, 90\%, 95\% and 99\% confidence levels. Black lines show the restframe $B$- and $V$-band extinction in magnitudes for a source at $z=1$, $A_B(z=1)$ and $A_V(z=1)$. \vspace{12pt}}
\label{chi2_Gra_81_single_nzconst}
\end{figure}

In Fig. \ref{chi2_Gra_81_single_nzconst}, constraints from the QSO colour analysis for the single size graphite dust model are shown. Regions (from yellow to red) indicate allowed dust models at 68\%, 90\%, 95\% and 99\% confidence levels. Black lines show the restframe $B$- and $V$-band extinction in magnitudes for a source at $z=1$, $A_B(z=1)$ and $A_V(z=1)$. Small size dust grains are more effectively constrained and the QSO colour analysis rules out any dimming $A_{B}(z=1) \gsim 0.05$ mag for grains with $a \lsim 0.2$ $\mu$m. As is evident from Fig. \ref{chi2_Gra_81_single_nzconst}, grains with radii $a \gsim 0.3\, \mu$m produce very little reddening for cosmological sources and $A_{B}(z=1) \approx A_{V}(z=1) \gsim 0.4$ mag can not be excluded. 

\begin{figure}
\includegraphics[width=84mm]{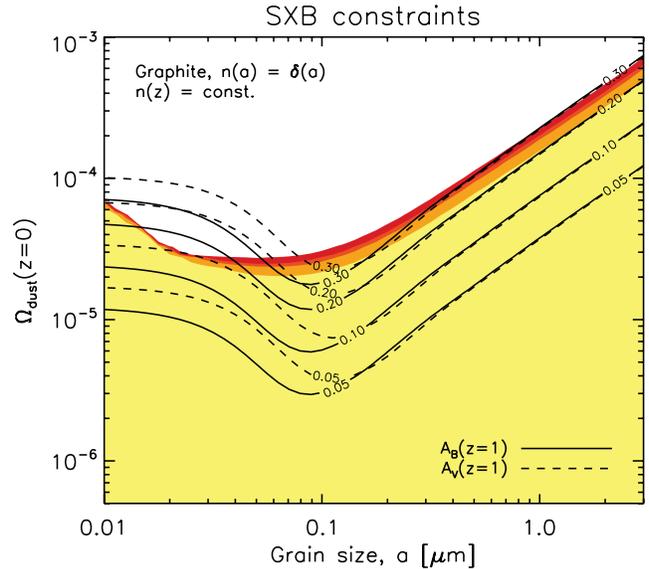}
\caption{Constraints from the unresolved Soft X-ray Background on dust models with single size graphite grains and a constant comoving number density. Regions (from yellow to red) indicate allowed dust models at 68\%, 90\%, 95\% and 99\% confidence levels. Black lines show the restframe $B$- and $V$-band extinction in magnitudes for a source at $z=1$, $A_B(z=1)$ and $A_V(z=1)$.}
\label{sxb_Gra_81_single_nzconst}
\end{figure}

\begin{figure}
\includegraphics[width=84mm]{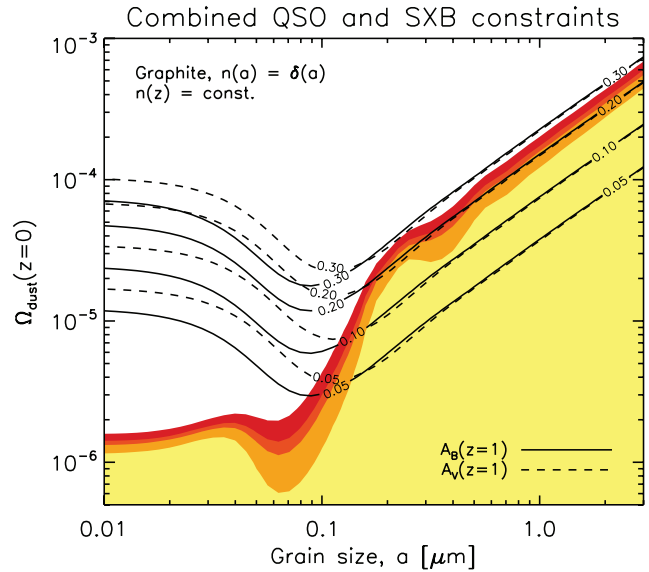}
\caption{Combined constraints from the QSO colour analysis and the unresolved Soft X-ray Background on dust models with single size silicate grains and a constant comoving number density. Regions (from yellow to red) indicate allowed dust models at 68\%, 90\%, 95\% and 99\% confidence levels. Black lines show the restframe $B$- and $V$-band extinction in magnitudes for a source at $z=1$, $A_B(z=1)$ and $A_V(z_=1)$.}
\label{combined_Gra_81_single_nzconst}
\end{figure}

\begin{figure}
\includegraphics[width=84mm]{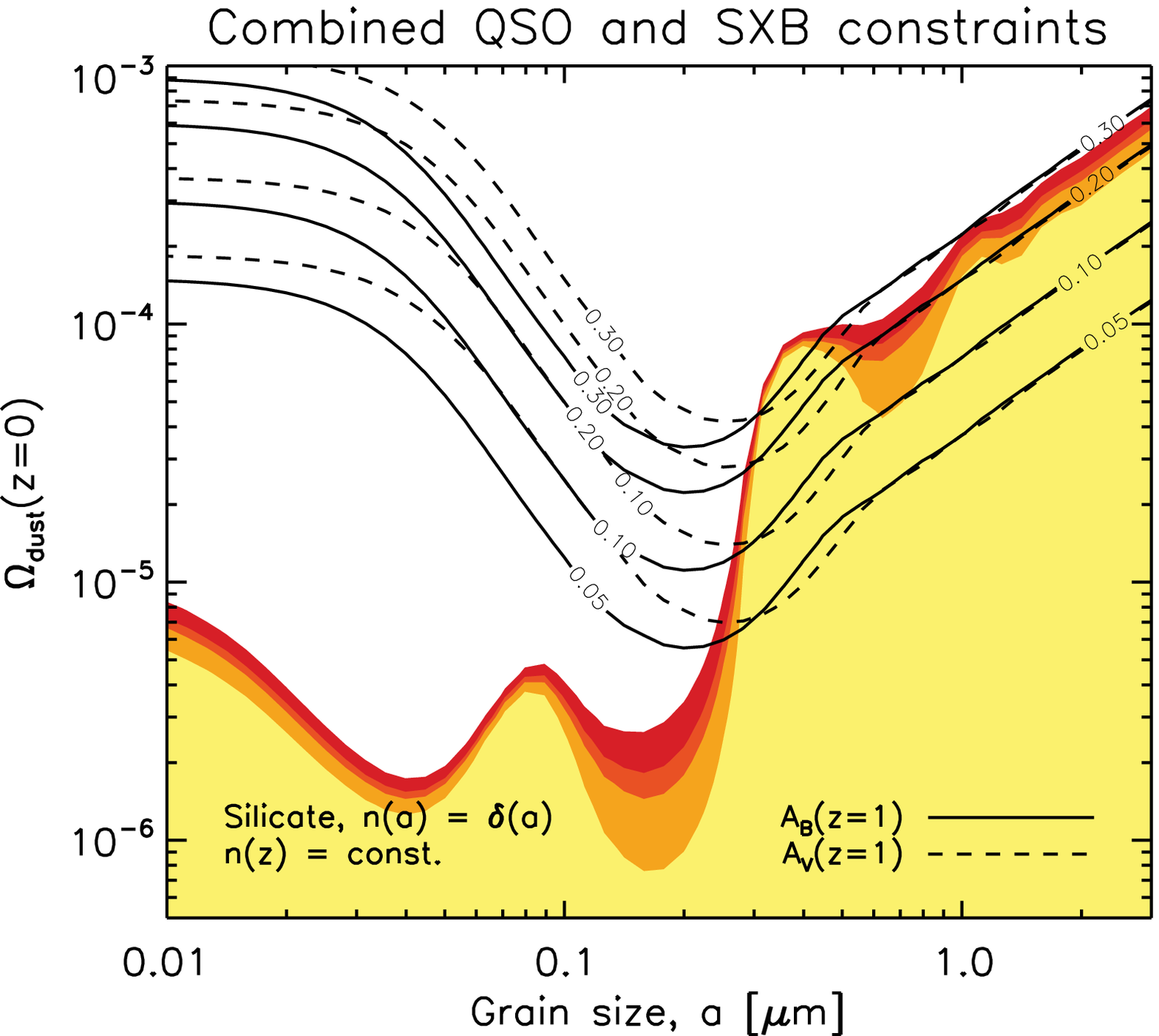}
\caption{Combined constraints from the QSO colour analysis and the unresolved Soft X-ray Background on dust models with single size silicate grains and a comoving number density proportional to the integrated star formation rate.
Regions (from yellow to red) indicate allowed dust models at 68\%, 90\%, 95\% and 99\% confidence levels. Black lines show the restframe $B$- and $V$-band extinction in magnitudes for a source at $z=1$, $A_B(z=1)$ and $A_V(z=1)$.}
\label{combined_suvSil_81_single_nzconst}
\end{figure}

Complementary to the QSO colour analysis, the SXB analysis (see Fig. \ref{sxb_Gra_81_single_nzconst}) is most effective for grains with sizes $a \gsim 0.2\,\mu$m, since for these grain sizes the scattering cross section for X-rays $\sigma_{\rm sca} \propto a^{4}$. Here, the unresolved SXB measurements constrain the the restframe $B$-band extinction, $A_{B}(z=1) \lsim 0.25$ mag at the 99\% confidence level. Combined constraints are shown in Fig. \ref{combined_Gra_81_single_nzconst}. For a dust population of single size silicate grains, results are both qualitatively and quantitatively similar as can be seen from the QSO and SXB combined constraints shown in Fig. \ref{combined_suvSil_81_single_nzconst}. 

Assuming the IGM consists solely of dust grains smaller than $a \sim 0.2 \, \mu$m with a constant comoving number density, we can constrain the dust density parameter $\Omega_{\rm dust} \lsim 10^{-6} - 10^{-5} (\rho_{\rm grain}/ 3{\rm \,g \,cm^{-3}})$, which is close to the astrophysically interesting levels implied by studies of stellar evolution and metallicity as a function of redshift. For grains larger than $a \sim 0.2 \, \mu$m, the constraints on $\Omega_{\rm dust} $ are degenerate with $a$. 

\subsection{Dust models with continuous grain size distributions}
Any realistic dust model will have a distribution of grain sizes. Here, we use an MRN model of separate populations of bare silicate and graphite grains with a power-law distribution of sizes where $dn/da \propto a^{-3.5}$. 

\begin{figure}
\includegraphics[width=84mm]{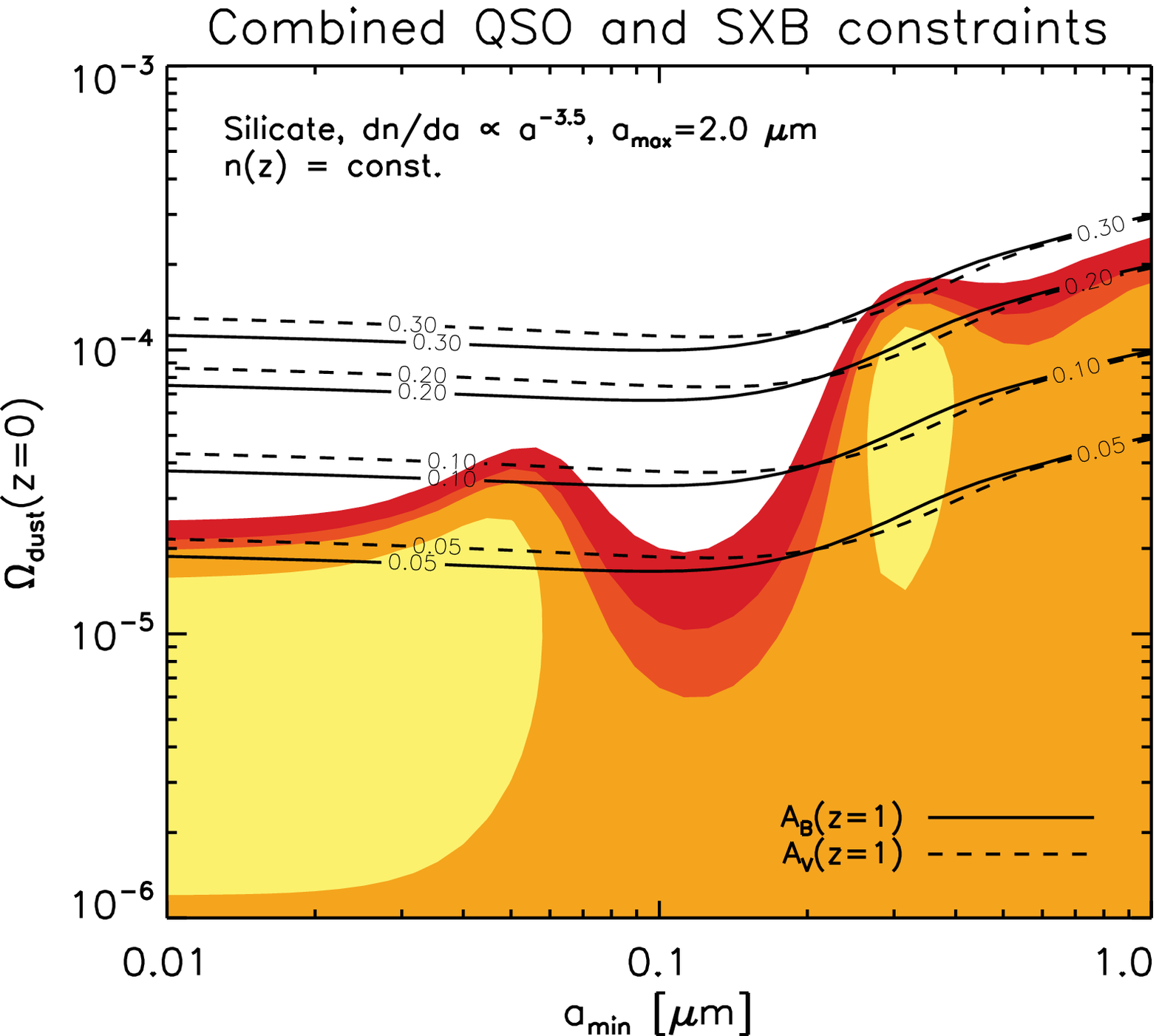}
\caption{Combined constraints from the QSO colour analysis and the unresolved Soft X-ray Background on dust models with silicate grains with a truncated MRN distribution of $0.01 < a_{\rm min} < 1$ $\mu$m and $a_{\rm max}=2 \, \mu$m and a constant comoving number density. Regions (from yellow to red) indicate allowed dust models at 68\%, 90\%, 95\% and 99\% confidence levels. Black lines show the restframe $B$- and $V$-band extinction in magnitudes for a source at $z=1$, $A_B(z=1)$ and $A_V(z=1)$.}
\label{combined_suvSil_81_MRNamin_nzconst}
\end{figure}

\begin{figure}
\includegraphics[width=84mm]{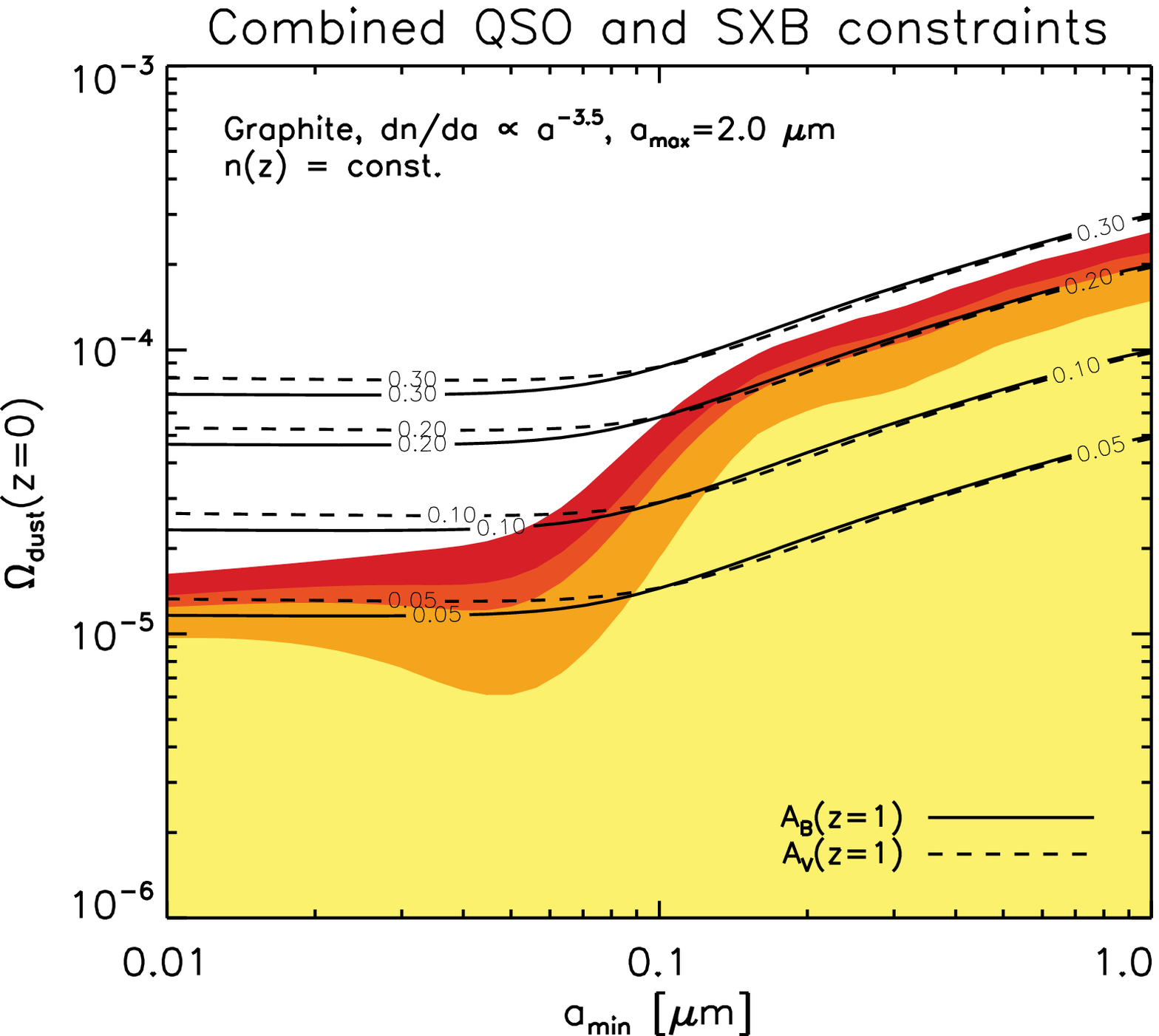}
\caption{Combined constraints from the QSO colour analysis and the unresolved Soft X-ray Background on dust models with graphite grains with a truncated MRN distribution of $0.01 < a_{\rm min} < 1$ $\mu$m and $a_{\rm max}=2$ $\mu$m and a constant comoving number density. Regions (from yellow to red) indicate allowed dust models at 68\%, 90\%, 95\% and 99\% confidence levels. Black lines show the restframe $B$- and $V$-band extinction in magnitudes for a source at $z=1$, $A_B(z=1)$ and $A_V(z=1)$.}
\label{combined_Gra_81_MRNamin_nzconst}
\end{figure}

First, we study the impact of the smallest grains in the MRN size distributions by varying $a_{\rm min}$ and keeping $a_{\rm max} = 2.0\, \mu$m. 
Figures \ref{combined_suvSil_81_MRNamin_nzconst} and \ref{combined_Gra_81_MRNamin_nzconst} show the combined QSO and SXB constraints on silicate and graphite grains with MRN size distributions and a constant comoving number density. These dust models all have a reddening parameter $6 \lsim \, R_{V} \lsim \infty$, representing very grey dust. Allowing dust grains $a \lsim 0.1$ $\mu$m, produces enough reddening at optical wavelengths that the QSO colours can limit the extinction in the $B$-band, $A_{B}(z=1) \lsim 0.10$. For models with $a_{\rm min} \gsim 0.1-0.3$ $\mu$m (completely grey dust) the SXB data helps to limit the extinction in the $B$-band to $A_{B}(z=1) \lsim 0.25$.

An interesting aspect of the MRN distribution is that the opacity is dominated by grains with small radii whereas the total dust mass is dominated by the grains with large radii. Thus, removing the very small grains can affect the opacity dramatically, without radically changing the total mass in dust. Therefore, for MRN distributions with $a_{\rm max} = 2.0\, \mu$m, the combined upper limits on the dust density are quite large, $\Omega_{\rm dust} \lsim 10^{-5} - 10^{-4} (\rho_{\rm grain} / 3 {\rm \,g \,cm^{-3}})$.

\begin{figure}
\includegraphics[width=84mm]{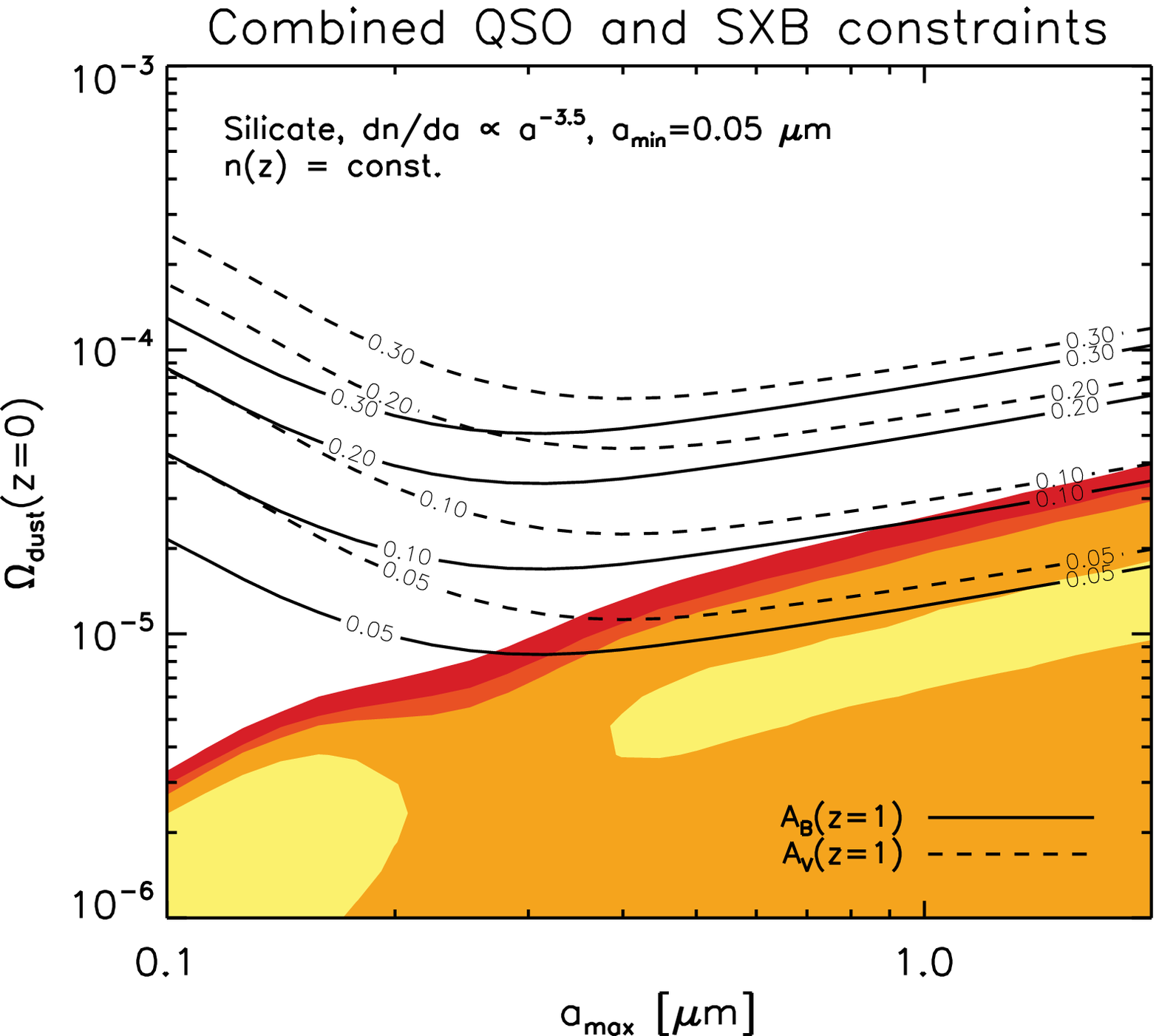}
\caption{Combined constraints from QSO colour analysis and the unresolved Soft X-ray Background on dust models with silicate grains with a truncated MRN distribution of sizes $a_{\rm min} = 0.05 \, \mu$m and $0.1 < a_{\rm max} < 2.0 \, \mu$m and a constant comoving number density. Regions (from red to yellow) indicate allowed dust models at 68\%, 90\%, 95\% and 99\% confidence levels. Black lines show the restframe $B$- and $V$-band extinction in magnitudes for a source at $z=1$, $A_B(z=1)$ and $A_V(z=1)$.}
\label{combined_suvSil_81_MRNamax_nzconst}
\end{figure}

\begin{figure}
\includegraphics[width=84mm]{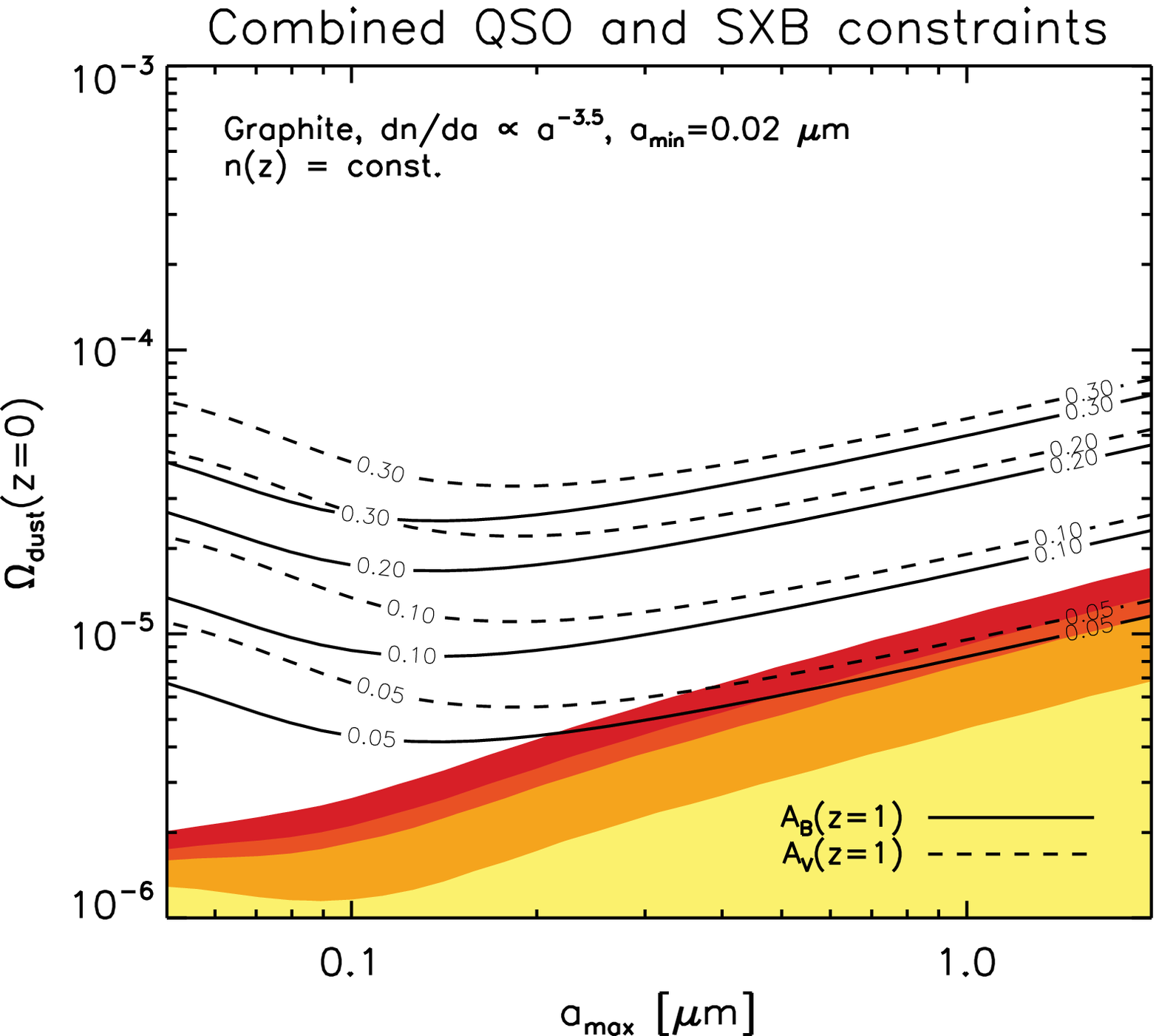}
\caption{Combined constraints from QSO colour analysis and the unresolved Soft X-ray Background on dust models with graphite grains with a truncated MRN distribution of sizes $a_{\rm min} = 0.02 \, \mu$m and $0.1 < a_{\rm max} < 2.0Ê\, \mu$m and a constant comoving number density. Regions (from yellow to red) indicate allowed dust models at 68\%, 90\%, 95\% and 99\% confidence levels. Black lines show the restframe $B$- and $V$-band extinction in magnitudes for a source at $z=1$, $A_B(z=1)$ and $A_V(z=1)$. }
\label{combined_Gra_81_MRNamax_nzconst}
\end{figure}

To relate to the detection of dust reddening of QSOs out to large distances around $z \sim 0.3$ galaxies  by \citet{menard2010a}, we focus on dust models with a reddening parameter close to their measured value of the reddening parameter, $R_V = 3.9 \pm 2.6$. Assuming a constant comoving dust density and extrapolating their result to $z=1$ yields a lower limit on the extinction in the restframe $B$-band of $A_{B}(z=1) \gsim 0.02$ magnitudes. 
If we keep $a_{\rm min}$ fixed at $0.05\, \mu$m (silicate grains, Fig. \ref{combined_suvSil_81_MRNamax_nzconst}) or at $0.02\mu$m (graphite grains, Fig.  \ref{combined_Gra_81_MRNamax_nzconst}) and vary the largest grain sizes  $0.1 < a_{\rm max} < 2\, \mu$m, the dust models span a range of $R_{V}$ between $\sim 1.3 - 7$. 
For these dust models, the combined QSO colour and SXB analysis puts an upper limit on the extinction in the restframe $B$-band, $A_{B}(z=1) \lsim 0.05 - 0.10$ mag and correspondingly limits the dust density parameter $\Omega_{\rm dust} \lsim 10^{-6} - 10^{-5} (\rho_{\rm grain} / 3 {\rm \,g \,cm^{-3}})$. 
\section{Summary and Discussion}
\begin{figure}
\includegraphics[width=84mm]{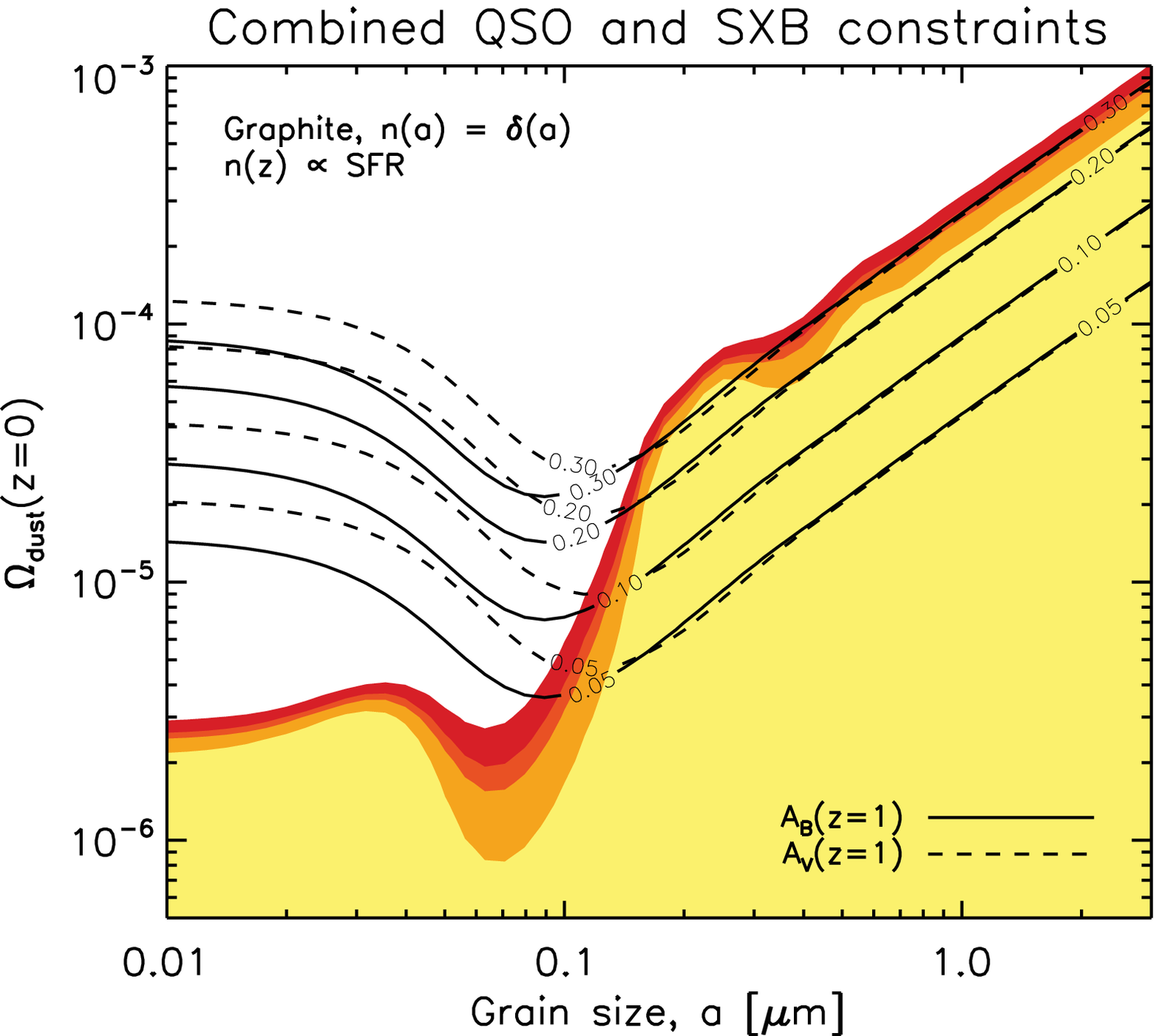}
\caption{Combined constraints from the QSO colour analysis and the unresolved Soft X-ray Background on dust models with single size graphite grains and a comoving number density proportional to the integrated star formation rate.
Regions (from yellow to red) indicate allowed dust models at 68\%, 90\%, 95\% and 99\% confidence levels. Black lines show the restframe $B$- and $V$-band extinction in magnitudes for a source at $z=1$, $A_B(z=1)$ and $A_V(z=1)$.}
\label{combined_Gra_81_single_nzsfr}
\end{figure}

In this paper, we studied the optical extinction and X-ray scattering effects of intergalactic dust grains. 
We find that dust distributions including graphite and silcate grains smaller than $a \lsim 0.1-0.3$ $\mu$m, produce enough reddening ($1.3 \lsim \, R_{V} \lsim 10$) to rule out extinction in the restframe $B$-band, $A_{B}(z=1) \gsim 0.05 - 0.10$ mag.
By combining the constraints from the QSO colour analysis with constraints from the unresolved SXB, we are able to rule out any systematic dimming in the restframe $B$-band for a source at redshift $z=1$ of $A_{B}(z=1) \gsim 0.25$ mag for a wide range of grain sizes (and consequently a large range of reddening parameters, $1.3 \lsim \, R_{V} \lsim \infty$). 
These results have been derived assuming a constant comoving dust density. In the case that the comoving dust density is proportional to the integrated star formation rate density, $n(z) \propto \int_{z}^{\infty} dz' \frac{\dot{\rho}_{\star}}{(1+z')\mathcal{E}(z')}$, the dust density decreases with redshift, making the universe increasingly transparent relative to the constant comoving dust density models. Results from these models are qualitatively similar to the models with constant comoving dust densities, except that the constraints on the induced dimming for sources at redshift $z=1$ (see Fig. \ref{combined_Gra_81_single_nzsfr}) are somewhat weaker, excluding $A_{B}(z=1) \gsim 0.05$ for small grains with $a \lsim 0.1\, \mu$m and $A_{B}(z=1) \gsim 0.35$ mag for large grains with $a \gsim 0.2\, \mu$m. However, for the same reasons, the upper limits on the dimming of sources at higher redshift ($z \gsim 2$) are stronger compared to the models with constant comoving dust density.

Since the error budget is dominated by systematic template uncertainties, the increase in the number of QSOs does not improve the limits from \citet{mortsell2003} and \citet{mortsell2005}, although the results are not directly comparable since the dust models probed are not equivalent. However, we regard the current error treatment as more realistic and also more effective in handling all possible colour combinations while accurately taking into account all possible colour and redshift correlations. 
In order to improve the current results from QSO colours (which are most effective in constraining the density of small size dust grains) it would be necessary to be able to decrease the systematic uncertainty of the continuum slope of the template spectrum. For larger dust grains, where the reddening effects are smaller and the dimming and density constraints from the unresolved SXB are weaker, limits could be improved by a better understanding of the different contributions to the unresolved SXB. Furthermore, improved constraints may be obtained by stacking observations of X-ray point sources. The absence of a single X-ray halo around a $z = 4.3$ QSO observed with Chandra, allowed \citet{petric2006} to place an upper limit on $\Omega_{\rm dust} \lsim 2 \times 10^{-6}$ assuming a constant comoving number density of dust grains with a characteristic grain size $a \sim 1\mu$m. 
This result is relaxed in \citet{corrales2012}, who find that $\Omega_{\rm dust} \lsim 10^{-5}$ for a wider range of dust models.
However, an increased number of sources will allow us either to measure or to improve the constraints on large dust grain densities, for which the corresponding limits on the dimming of cosmological sources are currently the weakest.

It should be noted that while smaller grains ($a \lsim 0.05\, \mu$m) may be either destroyed by sputtering or unable to travel far from formation sites as they are inefficiently pushed away by radiation pressure, large grains ($a \gsim 0.25\, \mu$m) may be too heavy and remain trapped in the gravitational field of the galaxy where they formed  \citep[]{aguirre1999b, aguirre2001, bianchi2005}. Combined with stellar evolution models \citep{fukugita2011}, the lower limits obtained in \citet{menard2010a} and the upper limits derived in this paper, we conclude that an MRN dust distribution of either silicate or graphite grains -- or possibly an admixture thereof -- with $a_{\rm min}\sim 0.05 \, \mu$m and  $a_{\rm max}\sim 0.25\, \mu$m and $\Omega_{\rm dust} \sim$ a few $\times 10^{-6} (\rho_{\rm grain} / 3 {\rm \,g \,cm^{-3}})$, constitute a viable intergalactic dust model.
\section*{Acknowledgments}
We would like to thank Ariel Goobar, Rahman Amanullah, Linda \"Ostman and Hugh Dickinson for useful discussions. We also wish to acknowledge the anonymous referee for helpful comments to the manuscript. EM acknowledge support for this study by the Swedish Research Council.
\bibliography{igdust}
 
\label{lastpage}
\end{document}